\documentclass[reprint,
superscriptaddress,
amsmath,amssymb,
aps,
pra,
]{revtex4-2}
\usepackage{graphicx}% Include figure files
\usepackage{dcolumn}% Align table columns on decimal point
\usepackage{bm}% bold math
\usepackage{dsfont}
\usepackage{physics}
\usepackage{color}
\begin{document} 
\title{ Resonant Cavity Modification of Ground State Chemical Kinetics}
\author{Lachlan P. Lindoy}%
\affiliation{Department of Chemistry, Columbia University, 3000 Broadway, New York, New York, 10027,  U.S.A}
\author{Arkajit Mandal}%
\affiliation{Department of Chemistry, Columbia University, 3000 Broadway, New York, New York, 10027,  U.S.A}
\author{David R. Reichman}
\email{drr2103@columbia.edu}
\affiliation{Department of Chemistry, Columbia University, 3000 Broadway, New York, New York, 10027,  U.S.A}
 
\begin{abstract}
 Recent experiments have suggested that ground state chemical kinetics can be suppressed or enhanced by coupling the vibrational degrees of freedom of a molecular system with a radiation mode inside an optical cavity. Experiments show that the chemical rate is strongly modified when the photon frequency is close to characteristic vibrational frequencies. The origin of this remarkable effect remains unknown. In this work, we develop an analytical rate theory for cavity-modified ground state chemical kinetics based on the Pollak-Grabert-H\"anggi rate theory. Unlike previous work, our theory covers the complete range of solvent friction values, from the energy-diffusion limited to the spatial-diffusion limited regimes. We show that the chemical reaction rate can either be enhanced or suppressed depending on the bath friction; when bath friction is weak chemical kinetics is enhanced as opposed to the case of strong bath friction, where chemical kinetics is suppressed. Further, we show that the photon frequency at which maximum modification of chemical rate is achieved is close to the reactant well, and hence resonant rate modification occurs. In the strong friction limit the {\it resonant} photon frequency is instead close to the barrier frequency, as obtained using the Grote-Hynes rate theory. Finally, we observe that the rate changes (as a function of photon frequency) are much sharper and more sizable in the weak friction limit than in the strong friction limit, and become increasingly sharp with decreasing well frequency. 
\end{abstract}

\maketitle
\section{Introduction} 
Vibrational polaritons~\cite{ThomasACID2016,ThomasS2019, NagarajanJACS2021}, quasi-particles formed by coupling molecular vibrations and radiation modes in an optical cavity, exhibit a wide range of exotic phenomena. A series of recent experiments~\cite{ThomasACID2016,ThomasS2019,LatherACID2019,VergauweACID2019,HiraiACID2020, AnoopNp2020,NagarajanJACS2021, LatherCS2022} have demonstrated that chemical kinetics can be enhanced~\cite{LatherACID2019, HiraiACID2020, LatherCS2022} or suppressed~\cite{ThomasACID2016, VergauweACID2019}, molecular bonds can be selectively broken~\cite{ThomasS2019}, and  selective crystallization can be achieved~\cite{HiraiCS2021} via the formation of vibrational polaritons. On the other hand, several studies~\cite{ImperatoreJCP2021, WiesehanJCP2021} have also reported possible discrepancies or inconsistencies in the interpretation of experiment purporting sizable kinetic effects. Thus, a rigorous theoretical understanding of the range of possible cavity-induced modifications to chemical kinetics is actively sought.  

Despite recent theoretical progress~\cite{LiNC2021, LiJPCL2021,TaoPNAS2020,GalegoPRX2019,JorgeJCP2020, YangJPCL2021,WangACSPh2021, MandalJCP2022, WangJPCL2022, FischerJCP2022, DuPRL2022, ChristianArxiv2022}, the fundamental theoretical understanding of cavity-modified ground state chemical reactivity remains inadequate. In short, the primary theoretical challenges regarding vibrational polaritonic chemistry include: (i)  explaining the  { resonance effect}, namely the observation that chemical kinetics is strongly modified when photon frequency is close to some vibrational frequency of the reactant molecule and (ii) explaining collective effects, namely the observation that cavity-modified chemical reactivity depends on the number of molecules coupled to the radiation mode.

Initial theoretical calculations employing simple transition state theory~\cite{LiJCP2020} (TST) concluded that coupling to the cavity does not modify the energy barrier and thus provides no change of chemical kinetics. Theoretical work that has employed multi-dimensional transition state theory, or equivalently the Grote-Hynes (GH) theory~\cite{GroteJCP1980},  has revealed a photon frequency dependent suppression of chemical kinetics at the single molecule level~\cite{LiNC2021, LiJPCL2021, FischerJCP2022}.  Further, the GH theory predicts that the maximum suppression of chemical kinetics is achieved when the photon frequency is close to the barrier frequency, and is independent of the reactant well frequency. This is in contrast to the experimental observations where the chemical kinetics is most strongly modified when the photon frequency is close to the reactant well frequency~\cite{ThomasACID2016, VergauweACID2019}.   

Quantum transition state theory (QTST)  predicts that the maximum suppression of chemical kinetics occurs when the photon frequency is between the barrier and the well frequency~\cite{YangJPCL2021}. When nuclear quantum effects are negligible, QTST provides the same result as the GH theory. Further, both GH theory and QTST only predict a mild suppression of chemical kinetics and a broad rate profile~\cite{LiNC2021, YangJPCL2021} as a function of photon frequency, in contrast to much larger and sharper rate profile typically observed in experiments~\cite{ThomasACID2016, VergauweACID2019,LatherCS2022}. On the other hand, computational studies~\cite{WangJPCL2022, SunArxiv2022} at the single molecular level have revealed enhancement of chemical kinetics~\cite{SunArxiv2022}, especially when the molecule-bath coupling is weak~\footnote{After the calculations in this work were completed, we became aware of Ref.~\citenum{SunArxiv2022}, where the low friction limit is explored numerically. Our theory is in agreement with these {\it in silico} experiments, and complements them by providing further analytical insights into the factors that govern the rate in this limit.}.

Note that all such reaction theories are either classical or semi-classical in nature, and do not address collective effects in a direct manner. Thus a theoretical explanation for such collective effects remains elusive. It has been argued  using GH theory that chemical kinetics can be suppressed when cavity radiation modes collectively couple to solvent vibrations that, in turn, are strongly coupled to the reaction coordinate~\cite{MandalJCP2022}.  In particular, theoretical work~\cite{YangJPCL2021} has suggested a   ``coherent mechanism" where the cavity-molecule coupling is scaled by $\sqrt{N}$ (where $N$ is number of molecules inside the cavity). 
However such an argument is heuristic and lacks rigorous theoretical justification. 

In this work, we develop a rigorous classical rate theory that provides the cavity-modified chemical rate for the full range of molecule-bath coupling strengths in the single molecule limit. Our work is based on the  Pollak-Grabert-H\"anggi rate theory (PGH)~\cite{PollakJCP1989} where the rate constant contains a depopulation factor that accounts for thermal activation in the energy-diffusion limited regime. In the weak molecule-bath coupling regime (i.e. the energy-diffusion limited regime), we show that coupling molecular systems to a radiation mode leads to an enhancement of chemical kinetics. This enhancement is greatest when the photon frequency is close to the reactant well frequency. Our results reveal that the rate profile as a function of photon frequency becomes sharper with decreasing reactant well frequency. We find that this resonance effect entirely originates from the depopulation factor, a term that is absent in GH theory. 

In the strong molecule-bath coupling limit the depopulation factor tends to unity and the PGH theory produces results identical to the GH theory. In this regime, we observe suppression of the chemical rate when coupling to a cavity, which corroborates previous theoretical calculations~\cite{LiNC2021, LiJPCL2021}. In this regime the maximum modification of the chemical rate occurs when the photon frequency is close to the barrier frequency as opposed to the reactant well frequency. The effect in the strong coupling limit is smaller than in the low friction limit. 

This letter is organized as follows: 
In sec.~\ref{theory} we outline the model. In sec.~\ref{rate-theory} we detail the theory of cavity-altered reaction rates. We present results in sec.~\ref{results}. In sec.~\ref{conclusion} we conclude. 

\section{Theory}
\subsection{Model Hamiltonian}\label{theory}
In this work we consider the Pauli-Fierz non-relativistic QED Hamiltonian in the dipole gauge and in the long-wavelength limit~\cite{Rubio2018JPB, Christian2020} with $\hbar = 1$ in atomic units (a.u.),
\begin{align}\label{Hlm}
\hat{H} &= \hat{H}_{\mathrm{m}} + \hat{H}_{\mathrm{c}} + \hat{H}_{\mathrm{b}} \nonumber \\
&= \frac{\hat{P}^2}{2} + V(\hat{Q}) \nonumber \\
&+  \frac{\hat{p}_\mathrm{c}^2}{2} + \frac{1}{2}\omega_\mathrm{c}^2\Bigg(\hat{q}_\mathrm{c} +  \sqrt{\frac{2}{\omega_\mathrm{c}^{3}}}\chi\cdot \hat{\mu}(\hat{Q})\Bigg)^2 \nonumber \\
&+ \sum_{i} \left[\frac{\hat{p}_i^2}{2} + \frac{1}{2}  \omega_i^2 \left(\hat{q_i} - \frac{c_i}{\omega_i^2} \hat{Q}\right)^2\right],
\end{align}
\begin{figure}
    \centering
    \includegraphics[width=0.9 \linewidth ]{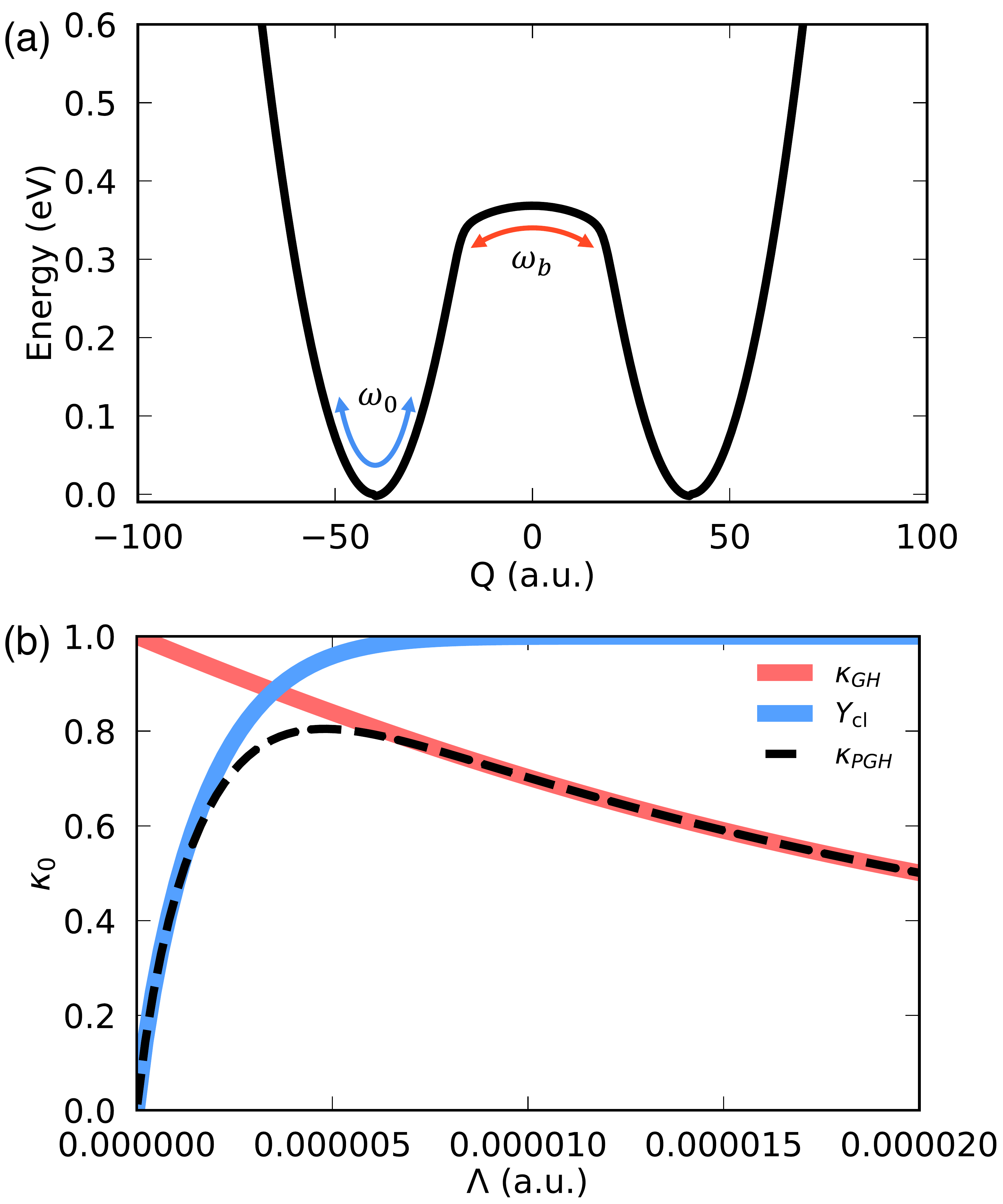}
    \caption{\small (a) Model potential energy surface $V(Q)$ as a function of the reaction coordinate Q. (b)  Grote-Hynes (GH) (red solid line) and Pollak-Grabert-H\"anggi (PGH) (black dashed line) transmission coefficients as a function of the bath reorganization energy $\Lambda$ when no cavity is present.}\label{fig:fig1}
\end{figure}
where  the last three lines describe the molecular Hamiltonian $\hat{H}_{\mathrm{m}} = \hat{P}^2/2 + V(\hat{Q})$, cavity Hamiltonian $\hat{H}_{\mathrm{c}} = \frac{\hat{p}_\mathrm{c}^2}{2} + \frac{1}{2}\omega_\mathrm{c}^2 (\hat{q}_\mathrm{c} +  \sqrt{2/\omega_\mathrm{c}^{3}} \chi\cdot \hat{\mu}(\hat{Q}) )^2$ and the molecular bath Hamiltonian $\hat{H}_{\mathrm{b}} = \sum_{i} [{\hat{p}_i^2/2} + \frac{1}{2}  \omega_i^2 (\hat{q_i} - \frac{c_i}{\omega_i^2} \hat{Q})^2]$, respectively.  
Here, $\hat{Q}$, $\hat{q}_\mathrm{c}$  and $\{ \hat{q}_{i}\}$  are position operators for the molecular reaction coordinate, the cavity radiation mode, and the bath vibrational modes, respectively. Note that in linear response theory, a harmonic bath may rigorously model an anharmonic solvent~\cite{BaderJCP1990,GeorgievskiiJCP1999}.  Here, $\omega_\mathrm{c}$ is the photon frequency, $\mu(\hat{Q})$ is the molecular dipole moment operator, and $\chi = \sqrt{\omega_{c}/2\epsilon V}$ characterizes light-matter coupling strength, where $\epsilon $ and $V$ is the permittivity of the medium placed between two cavity mirrors and the quantization volume, respectively. Lastly, $\omega_{i}$  is the $i$th  bath mode frequency for a vibration  that couples to the reaction coordinate with a coupling constant $c_{i}$. 

We   consider  an adiabatic ground state chemical reaction described by a symmetric double well potential 
$V(\hat{Q})$ (shown in Fig.~\ref{fig:fig1}a). Modification of our approach to the asymmetric case is trivial. The ground adiabatic potential energy surface $V(\hat{Q})$ is characterized by a reactant well frequency $\omega_{0}$  and a barrier frequency $\omega_{b}$ such that,

\begin{align}
V({Q}\approx Q_{r}) &\approx ~~\frac{1}{2}\omega_{0}^{2} ({Q} - Q_{r})^{2}, \nonumber\\
V({Q}\approx Q_{b}) &\approx -\frac{1}{2}\omega_{b}^{2} ({Q} - Q_{b})^{2} + E_{b},
\end{align} 
where $Q_{r}$ and $Q_{b}$ are the position of reactant and the barrier, respectively, and $E_{b}$ is the barrier height. Details of the model molecular system are provided in the Supporting Information (SI). 

The cavity Hamiltonian $\hat{H}_{\mathrm{c}}$ used in this work describes the interaction between a single cavity mode and the molecular system through the ground state molecular permanent dipole moment $\hat{\mu}(Q)$. For simplicity, we assume a linear dipole $\hat{\mu}(Q) = \mu_{0}\cdot Q$. Noting that $\chi \propto \omega_{\mathrm{c}}$, we write   $\hat{H}_{\mathrm{c}}$ as
\begin{align}
\hat{H}_{\mathrm{c}} = \frac{\hat{p}_\mathrm{c}^2}{2} + \frac{1}{2}\omega_\mathrm{c}^2\Bigg(\hat{q}_\mathrm{c} +  \frac{\eta_{\mathrm{c}} }{\sqrt{\omega_{c}}}\cdot \hat{Q}\Bigg)^2, 
\end{align} 

where $\eta_{\mathrm{c}} = \sqrt{2} \mu_{0} \frac{\chi}{\omega_\mathrm{c}}$ is independent of the photon frequency $\omega_\mathrm{c}$. 

The vibrational frequencies $\omega_{i}$ and couplings $c_{i}$ in the molecular bath Hamiltonian  $\hat{H}_{b}$ are sampled from a spectral density defined as
\begin{equation}
J_{b}(\omega) = \frac{\pi}{2}\sum_i \frac{c_i^2}{ \omega_i} \delta(\omega-\omega_i), 
\end{equation}
which here is taken to be a standard Debye spectral density often used to model chemical solvents~\cite{MukamelBook}
\begin{equation}
    J_{b}(\omega) = \frac{2\Lambda \Omega_c\omega}{\omega^2+\Omega_c^2}, 
\end{equation}
with cutoff frequency $
\Omega_c = 0.3$ eV, and reorganization energy $\Lambda$.  All calculations have been run at $T = 300$ K.

\subsection{Rate Theory}\label{rate-theory}
Following the Pollak-Grabert-H\"anggi theory (PGH)~\cite{PollakJCP1986}, the chemical rate constant for the light-matter Hamiltonian in Eq.~\ref{Hlm} in the classical limit ($\hat{H} \rightarrow \mathcal{H}(Q,q_\mathrm{c}, \{q_{i}\})$) can be expressed as,

\begin{equation}\label{PGH}
k = Y_\mathrm{cl} \cdot \kappa_{\mathrm{GH}} \cdot k_{\mathrm{TST}} =    \kappa_{\mathrm{PGH}} \cdot 
\frac{\omega_{0}}{2\pi} e^{-\beta E_{b}},
\end{equation}
where $k_{\mathrm{TST}}= \frac{\omega_{0}}{2\pi} e^{-\beta E_{b}}$ is the simple transition state theory rate constant, and  $\kappa_{\mathrm{GH}}$ is the transmission coefficient within the GH theory~\cite{PeterRMP1990, EyringJCP1935, SlaterJCP1956, GroteJCP1980}. Most importantly for this work, $Y_\mathrm{cl}$ is the classical depopulation factor which determines the total transmission coefficient $\kappa_{\mathrm{PGH}} = Y_\mathrm{cl}\cdot \kappa_{\mathrm{GH}}$~\cite{PollakJCP1986}. 

The GH transmission coefficient $\kappa_{\mathrm{GH}}$ is given by

\begin{equation}
   \kappa_{\mathrm{GH}} = \frac{\lambda_b}{\omega_b},
    \label{eq:kappa_GH}
\end{equation}
where $\lambda_b$ is the frequency of the unstable normal mode of the molecule-cavity-bath hybrid system which can be obtained from the GH relation~\cite{GroteJCP1980}
\begin{equation}
    \lambda_b = \frac{\omega_b^2}{\lambda_b + \gamma(\lambda_b)}.
\end{equation}
Here $\gamma(z)$ is the Laplace transform of the time-dependent friction, which can be obtained from the total spectral density~\cite{PollakJCP1989},
\begin{equation}\label{laplaceFriction}
    \gamma(z) = \frac{2}{\pi}\int_0^\infty \frac{J(\omega)}{\omega} \frac{z}{\omega^2+z^2}.
\end{equation}

Note that the total spectral density is $J(\omega) = J_{b}(\omega) + \frac{\pi}{2}  {\eta_\mathrm{c}^2 \omega_\mathrm{c}^{2}}  \delta(\omega-\omega_\mathrm{c})$ such that  the cavity mode is effectively treated as an additional vibrational mode. 

For a double-well potential, the depopulation factor, $Y_\mathrm{cl}$, can be expressed as~\cite{MelnikovJCP1986,TopalerJCP1994}
\begin{equation}
    Y_\mathrm{cl} = \frac{Y_\mathrm{m}(\delta_L) Y_\mathrm{m}(\delta_R)}{Y_\mathrm{m}(\delta_L+\delta_R)}, \label{eq:depopulation_factor}
\end{equation}
where 
$Y_\mathrm{m}$ is the classical depopulation factor for the escape from a metastable state and is given by~\cite{MelnikovJCP1986,PollakJCP1989, RipsPRA1990}
\begin{equation}
    Y_\mathrm{m}(\delta)= \exp\Bigg[\frac{1}{\pi}\int_{-\infty}^{\infty} \frac{dy}{1+y^2} \mathrm{ln}\Big(1 - e^{ -\frac{\delta}{4}(1+y^2)}\Big)\Bigg].
\end{equation}
In Eq. \ref{eq:depopulation_factor}, $\delta_L$ and $\delta_R$ are the average (dimensionless) energy loss from the unstable mode when the system returns to the barrier, associated with the left and right wells, respectively, and can be obtained from~\cite{TopalerJCP1994}
\begin{equation}
    \delta = \frac{\beta}{2\pi}\int_{-\infty}^\infty \mathrm{Re}\left[K(iz)\right] f(z) dz. \label{eq:energy_loss}
\end{equation}
Here
\begin{equation}
    K(z) = \frac{1}{u_{00}^2} \frac{z}{z^2 + z \gamma(z) - \omega_b^2} - \frac{z}{z^2-\lambda_0^2}
\end{equation} 
is the Laplace transform of the classical dissipation kernel, where $u_{00}$ is the coefficient of the unstable mode in the normal mode expansion of the mass-weighted reaction coordinate, which can be obtained from the spectral density as~\cite{PollakJCP1989,TopalerJCP1994}
\begin{equation}
    u_{00}^2 = \left[1 + \frac{2}{\pi}\int_0^\infty \frac{J(\omega)\omega}{\left(\omega^2+\lambda_0^2\right)^2} \right]^{-1}.
\end{equation}
In Eq. \ref{eq:energy_loss}, 
$f(z) = \left|\int_{-\infty}^{\infty} e^{izt} F(t) dt\right|^2$
is obtained from the time-dependent force
\begin{equation}
    F(t) = -{u_{00}}  \frac{\mathrm{d}V_\mathrm{NL}}{\mathrm{d}Q}\Bigg|_{Q= {u_{00}} \mathcal{Q}(t)}
\end{equation}
which arises from the classical trajectory of the unstable mode at the barrier energy, $\mathcal{Q}(t)$, where $V_\mathrm{NL}(Q) = V(Q) + \frac{1}{2} \omega_b^2 Q^2$
accounts for the non-linear contributions to the force at the top of the barrier~\cite{PollakJCP1989, RipsPRA1990,TopalerJCP1994}.
The unstable mode trajectory is obtained as the solution of the equation of motion,
\begin{equation}
    \ddot{\mathcal{Q}}(t) - \lambda_0^2 \mathcal{Q}(t) = F(t),
\end{equation}
starting at the top of the barrier and returning to the barrier with period $\tau\rightarrow\infty$.
\begin{figure}
    \centering
    \includegraphics[width=0.9\linewidth]{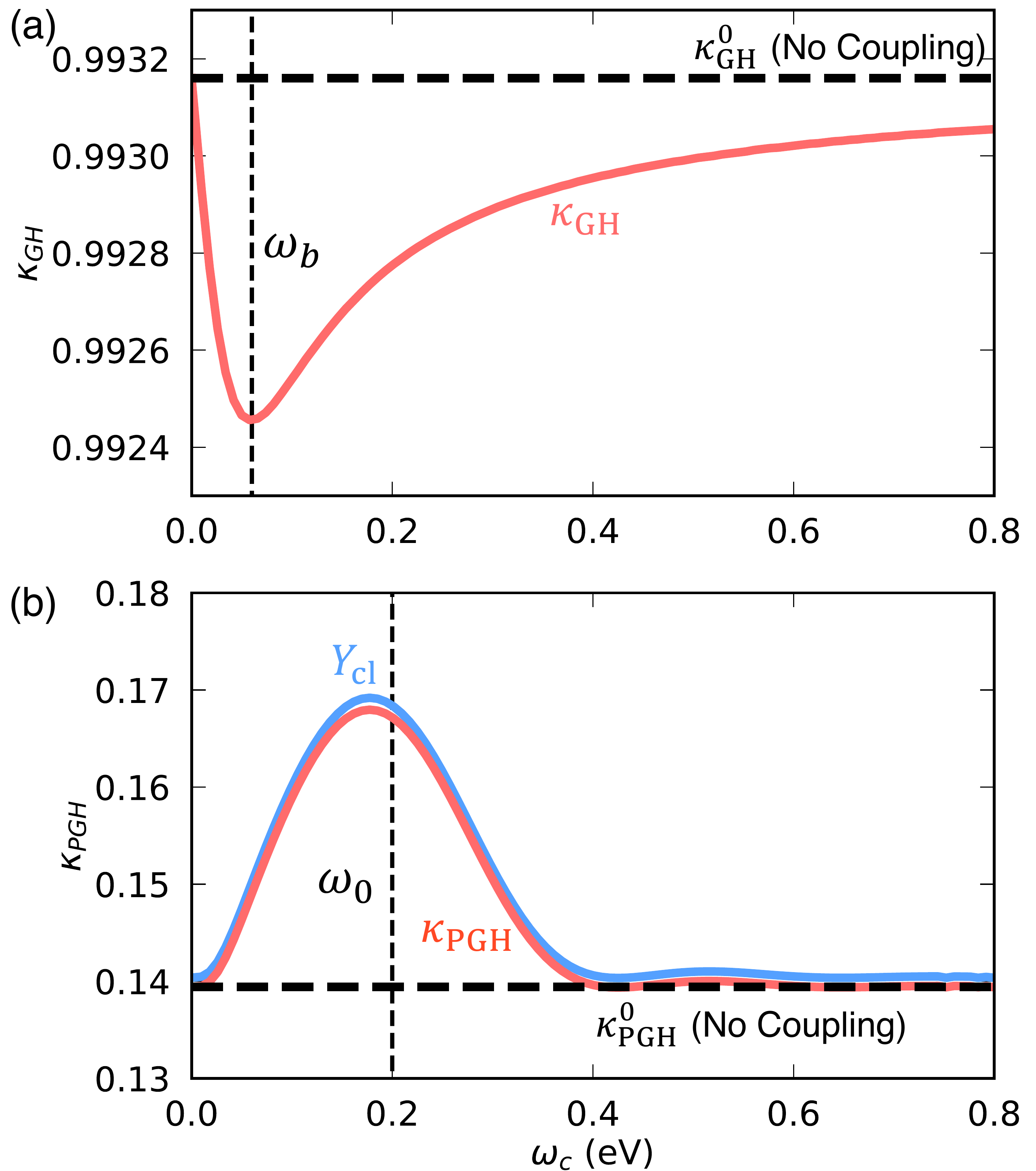}
    \caption{\small Dependence of the (a) Grote-Hynes   and (b) Pollak-Grabert-H\"anggi transmission coefficients on the cavity frequency in the energy-diffusion limited regime ($\Lambda = 2\times10^{-7}$ a.u.) with $\eta = 0.0025$ a.u.  and a reaction coordinate frequency $\omega_0 = 0.2$ eV).  Note the dramatically different scales for the transmission coefficients.}
    \label{fig:fig2}
\end{figure}
 
When the system-bath coupling is strong $Y_\mathrm{cl} \rightarrow 1$, and the chemical rate constant becomes $k \approx \kappa_\mathrm{GH}\cdot k_\mathrm{TST}$, such that the PGH theory reduces to the GH theory.  In this regime, the strong system-bath coupling leads to rapid thermalization of the energy in the unstable mode, a central assumption in GH theory~\cite{PollakJCP1986}.  In the weak system-bath coupling regime, this no longer holds and it becomes necessary to explicitly account for the exchange of energy between the unstable and stable modes.  Within the PGH theory, the effect of this energy transfer on the transmission coefficient is accounted for by the depopulation factor, $Y_\mathrm{cl}$.  In the weak coupling regime $Y_\mathrm{cl} \ll 1$, and the overall rate becomes incredibly sensitive to the average energy loss, $\delta$,  that characterizes the transfer of energy between the unstable and stable modes. PGH theory extends the GH theory to capture the Kramers turnover between the energy-diffusion limited regime and the spatial-diffusion limited (high-friction) regimes.  We note that the PGH theory is generally in semi-quantitative agreement with direct numerical simulation across all values of friction~\cite{PollakJCP1986}.

The general behavior of our system when decoupled from the cavity ($\hat{H} \rightarrow \hat{H} - \hat{H}_{c}$) is shown in Fig.~\ref{fig:fig1}b. In Fig.~\ref{fig:fig1}b we compute the transmission coefficient $\kappa = k/k_\mathrm{TST}$ from GH theory ($\kappa_\mathrm{GH}$) and PGH theory ($\kappa_\mathrm{PGH} = Y_\mathrm{cl}\cdot \kappa_\mathrm{GH}$). The red solid line represents the GH transmission coefficient $\kappa_\mathrm{GH}$ as a function of the bath reorganization energy $\Lambda$ compared to the PGH transmission coefficient $\kappa_\mathrm{PGH}$, represented by the black dashed line. The PGH theory shows the characteristic Kramers turnover~\cite{PeterRMP1990}, in contrast to the red solid line that monotonically decreases with increasing $\Lambda$ and deviates from the complete theory below $\Lambda \sim 5\times 10^{-6}$ a.u. Note that for our chosen form of the potential and parameters, the GH theory is accurate just past the turnover point. For other parameters it may be the case that the GH theory is only asymptotically accurate for large friction. The PGH theory can indeed capture this range of friction as well. 

\section{ Results and discussion}\label{results} Fig.~\ref{fig:fig2} presents the  transmission coefficient  for a molecule-cavity-bath hybrid system described in Eq.~\ref{Hlm} computed from the GH and PGH theories. In Fig.~\ref{fig:fig2} we vary the photon frequency while keeping $\eta_{c} = 0.0025$ a.u. a constant and employ a small bath reorganization energy $\Lambda = 2 \times 10^{-7}$ a.u. 
 
 In Fig.~\ref{fig:fig2}a the photon frequency, $\omega_\mathrm{c}$, dependence of the GH transmission coefficient, $\kappa_\mathrm{GH}$, is presented. Note that the GH theory does not depend on $\omega_{0}$ and only depends on $\omega_{b}$~\cite{LiNC2021}, as evident from Eqs.~\ref{eq:kappa_GH}-\ref{laplaceFriction}. Overall, $\kappa_\mathrm{GH}$ shows suppression of the chemical rate, in comparison to the no coupling ($\eta_{c} = 0$) scenario represented by the black solid line. However, for the coupling values chosen here the suppression is very weak. Additionally, $\kappa_\mathrm{GH}$ exhibits a minimum when the photon frequency is close to $\omega_{b}$, as has been shown in recent work~\cite{LiNC2021,YangJPCL2021, LiJPCL2021}. Note that $\kappa_\mathrm{GH}$ exhibits very broad suppression and very slowly approaches the zero coupling results over a frequency range  of $\sim 2$ eV on the scale of $\kappa_\mathrm{GH}$ presented here.
 \begin{figure}
    \centering
    \includegraphics[width=0.98\linewidth]{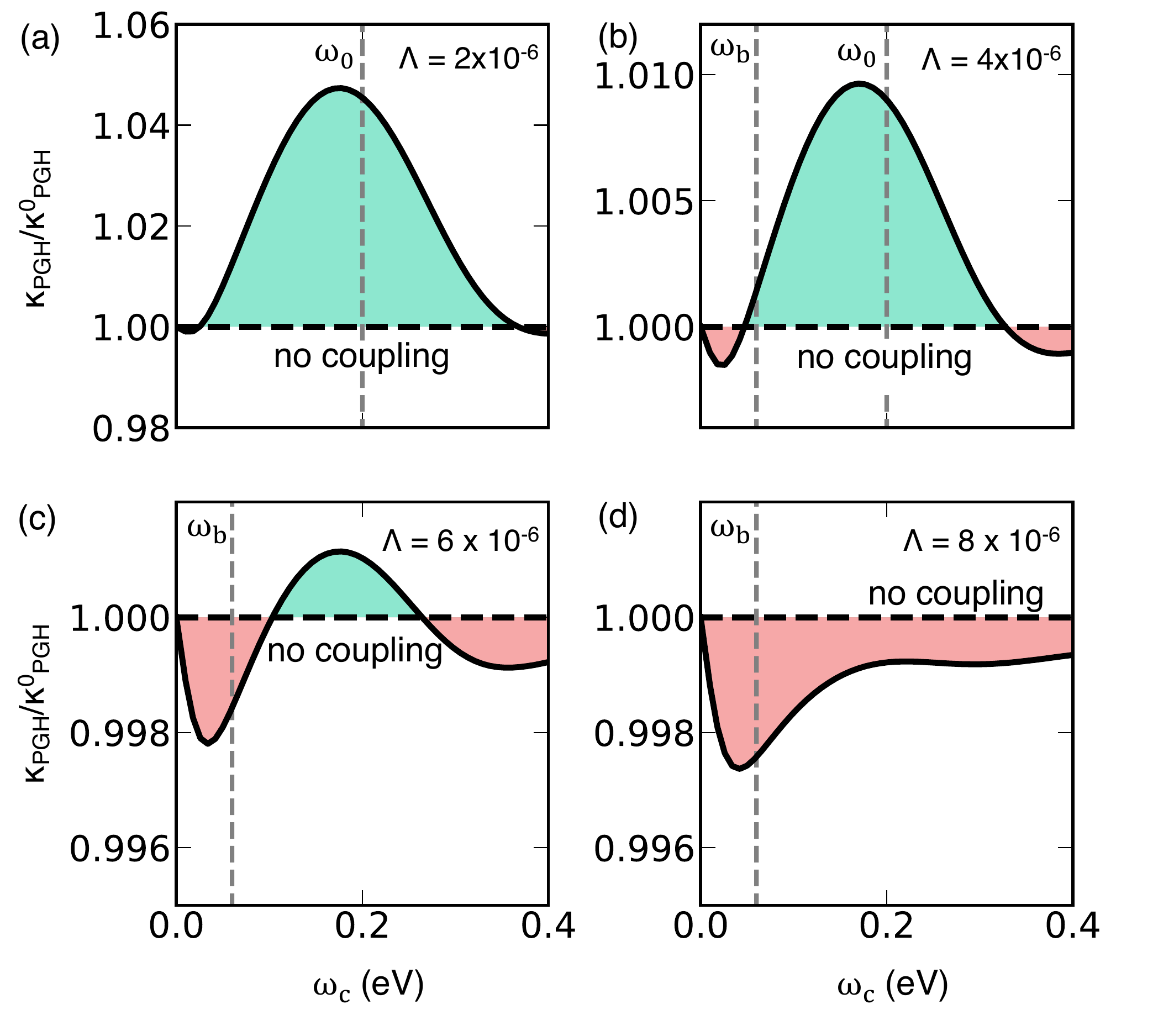}
    \caption{\small Cavity-modified chemical rate computed using Pollak-Grabert-H\"anggi (PGH) theory at various solvent friction values, (a)  $\Lambda = 2\times10^{-6}$ a.u., (b) $\Lambda = 4\times 10^{-6}$ a.u., (c) $\Lambda = 6\times 10^{-6}$ a.u. and (d) $\Lambda = 8\times 10^{-6}$ a.u. and with a constant light-matter coupling $\eta_\mathrm{c} = 0.005$ a.u., with green and red shading representing enhancement or suppression, respectively. $\kappa^{0}_\mathrm{PGH}$ is the total transmission coefficient when no cavity is present. }
    \label{fig:fig3}
\end{figure}

 Fig.~\ref{fig:fig2}b presents the  PGH transmission coefficient $\kappa_\mathrm{PGH}$ for the cavity-modified rate as a function of $\omega_\mathrm{c}$ with the same $\eta_{c}$ as in Fig.~\ref{fig:fig2}a.  Importantly, $\kappa_\mathrm{PGH}$ displays quantitatively as well as qualitatively different rate behavior.  First, in Fig.~\ref{fig:fig2}b we observe cavity-mediated enhancement in contrast to the suppression  predicted within GH theory. Second, the maximum cavity modification occurs when photon frequency is close to the reactant well frequency, that is $\omega_\mathrm{c} \approx \omega_{0}$. This qualitative feature of the PGH theory matches experimental observations, where maximum cavity modification is also achieved when $\omega_\mathrm{c} \approx \omega_{0}$~\cite{LatherCS2022, LatherACID2019, HiraiACID2020}. However, we emphasize that in this work we have considered only a single molecule coupled to a cavity mode. Thus caution must be exercised when comparing to the experiments where a large ensemble of molecular vibrations are collectively coupled to  the cavity.  Third, the extent of cavity modification predicted within the PGH theory is much larger than that predicted in the GH theory. In Fig.~\ref{fig:fig2}a the modification of  chemical rate is at most $0.07\%$, whereas the chemical rate is modified by $\approx 21\%$ in Fig.~\ref{fig:fig2}b when using the PGH theory. Lastly, the width of the enhancement of the rate as a function of frequency is much narrower than that of suppression in the high friction limit. The PGH transmission coefficient 
  in this  low molecular bath friction regime is dominated by the depopulation factor $Y_\mathrm{cl}$ (blue solid line in Fig.~\ref{fig:fig2}b) such that $\kappa_\mathrm{PGH}$ primarily inherits its shape from $Y_\mathrm{cl}$.

 The light-matter coupling modulates the depopulation factor $Y_\mathrm{cl}$ and the behavior of $\kappa_\mathrm{GH}$ (note that $\kappa_\mathrm{PGH} = Y_\mathrm{cl}\cdot \kappa_\mathrm{GH}$) in two different ways. The light-matter coupling significantly increases the energy loss, $\delta$, accounting for transfer of energy between the stable and unstable modes, increasing $Y_\mathrm{cl}$, which leads to an enhancement of the chemical reaction rate. At the same time, the light-matter coupling decreases $\kappa_\mathrm{GH}$, which leads to suppression of the reaction rate. The overall modification of the reaction rate is a result of an interplay of these two effects. The relative extent to which both of these terms ($Y_\mathrm{cl}$ and $\kappa_\mathrm{GH}$) are modified depends on the bath friction. Consequently, the deviation of the PGH theory from the GH theory also depends on the bath friction. We illustrate this in Fig.~\ref{fig:fig3}, where we consider four different bath reorganization energy values $\Lambda$ at a fixed light-matter coupling strength $\eta_\mathrm{c} = 0.005$ a.u.  

\begin{figure}
    \centering
    \includegraphics[width=0.9\columnwidth]{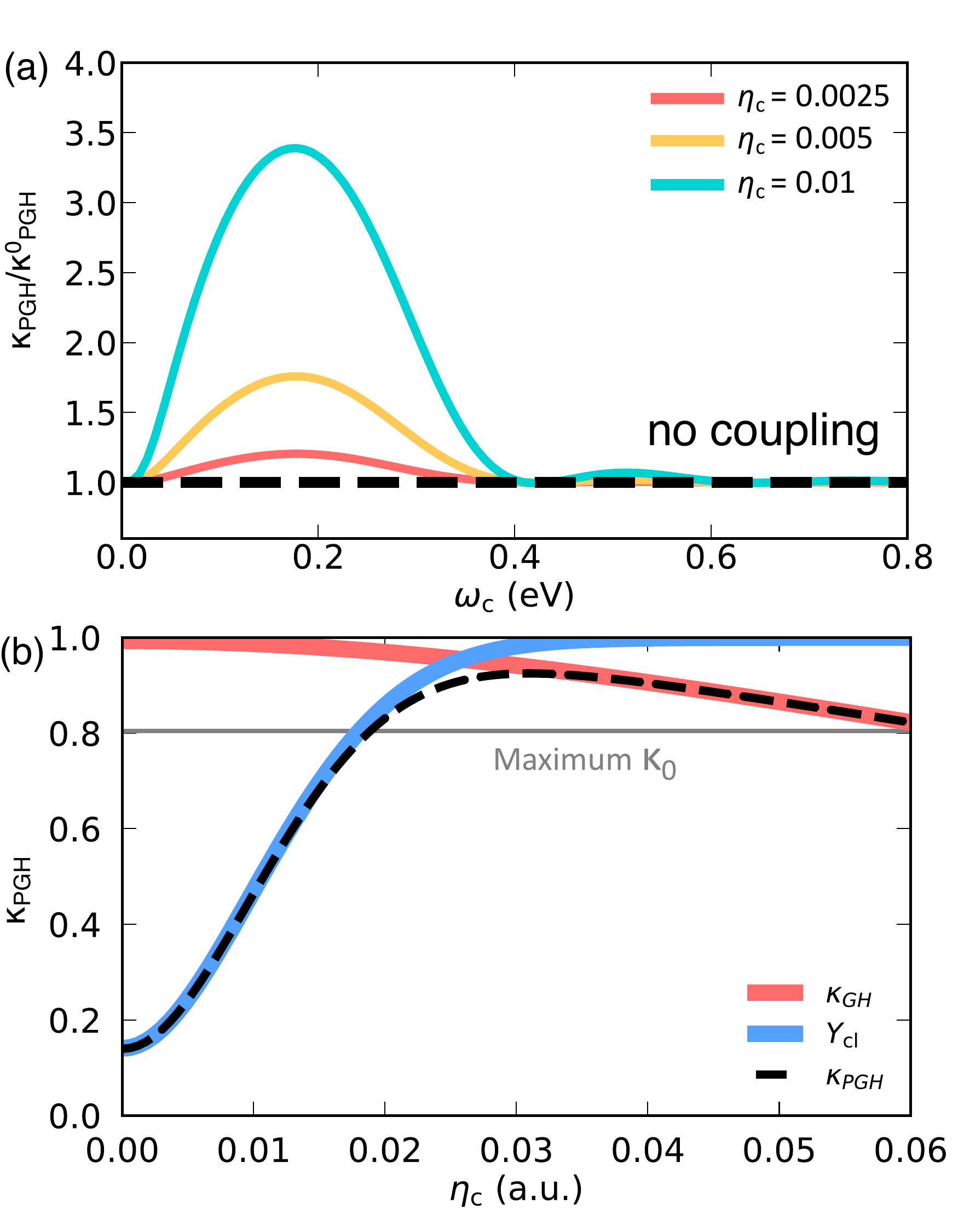}
    \caption{Dependence of the Pollak-Grabert-H\"anggi (PGH) transmission coefficients on the cavity frequency in the energy-diffusion limited regime ($\Lambda = 2\times10^{-7}$ a.u.) (a) for different values of the light-matter coupling $\eta_{c}$ as function of photon frequency $\omega_{c}$ and (b) as a function of $\eta_{c}$ with $\omega_{c} =\omega_{0} = 0.2$ eV. }
    \label{fig:fig4}
\end{figure}

Figs.~\ref{fig:fig3}a-d present normalized transmission coefficient $\kappa_\mathrm{PGH}/\kappa^0_\mathrm{PGH}$, where $\kappa^0_\mathrm{PGH}$ is transmission coefficient in the no coupling scenario ($\eta_\mathrm{c} = 0$) computed using the PGH rate theory. In Figs.~\ref{fig:fig3}a-b, we set $\Lambda = 2 \times 10^{-6}$ a.u. and $4 \times 10^{-6}$ a.u. respectively, such that the bath friction lies in the energy-diffusion limited regime. In this limit, just as in Fig.~\ref{fig:fig2}b, light-matter interactions lead mostly to enhancement of chemical rate (indicated by the green shaded region) which is maximized when the photon frequency is resonant with  $\omega_{0}$.  We note that there is a small deviation between $\omega_{0}$ and the $\omega_{c}$ at which the $Y_\mathrm{cl}$ is maximum. This is due to the non-linearity of $V(Q)$, such that it is only approximately a harmonic potential of frequency $\omega_{0}$ near the reactant well. Meanwhile we also observe a small amount of suppression in the transmission coefficient in Fig.~\ref{fig:fig3}b (indicated by the red shaded region) when $\omega_\mathrm{c}$ is far from $\omega_\mathrm{0}$ due to the contribution of the cavity-modified $k_\mathrm{GH}$. 

In Fig.~\ref{fig:fig3}c, we consider a relatively high bath friction value $\Lambda = 6 \times 10^{-6}$ a.u., which lies near the maximum of $\kappa_\mathrm{PGH}$ in Fig.\ref{fig:fig1}b. Interestingly, in this case,  the interplay between $Y_\mathrm{cl}$ and $\kappa_\mathrm{GH}$ becomes prominent, and we observe both suppression as well as enhancement of the chemical reaction rate. Enhancement occurs when $\omega_\mathrm{c}\approx \omega_{0}$ and suppression when $\omega_\mathrm{c}$ is far from $\omega_{0}$. 

In Fig.~\ref{fig:fig3}d we consider stronger bath friction $\Lambda = 8 \times 10^{-6}$ a.u.. In this regime,
$Y_\mathrm{cl} \approx$ 1 in the absence of the cavity ($\eta_\mathrm{c} = 0$). Thus turning on the cavity coupling, which effectively increases overall friction in Eq.~\ref{laplaceFriction}, does not lead to a noticeable increase in $Y_\mathrm{cl}$ since it is bounded by 1 ($0 \leq Y_\mathrm{cl} \leq 1$). As a result, $\kappa_\mathrm{PGH} \approx \kappa_\mathrm{GH}$, and one observes the characteristic chemical suppression as predicted by the GH theory~\cite{LiNC2021}.  

Overall, the cavity mode can be regarded as an addition bath degree of freedom which effectively increases the bath friction in Eq.~\ref{laplaceFriction}. 
\emph{ Thus, the light-matter coupling strength can be regarded as a control knob that can tune the effective environmental friction of the molecular system.} We illustrate this in Fig.~\ref{fig:fig4}.    

 In Fig.~\ref{fig:fig4}, we analyze the effect of light-matter coupling on cavity-modified chemical kinetics. Fig.~\ref{fig:fig4}a present the normalized transmission coefficient $\kappa_\mathrm{PGH}$ as a function of $\omega_\mathrm{c}$ at three different values of $\eta_\mathrm{c}$. Here we use $\Lambda = 2 \times 10^{-7}$ a.u. which lies in the energy-diffusion limited regime. As expected, we observe that increasing the light-matter coupling $\eta_\mathrm{c}$ leads to an increase in the chemical reaction rate. Similar to the original Kramers turnover behavior, increasing $\eta_\mathrm{c}$ further leads to a turnover of the chemical rate, where increasing  $\eta_\mathrm{c}$ leads to a decrease in the chemical reaction rate. This is shown in Fig.~\ref{fig:fig4}b.  
 
 In Fig.~\ref{fig:fig4}b we present the 
 cavity-modified transmission coefficient $\kappa_\mathrm{PGH}$ as a function  $\eta_\mathrm{c}$ and at a constant cavity photon frequency $\omega_\mathrm{c}  = \omega_{0}$. For small  light-matter coupling strengths ($\eta_\mathrm{c} < 0.02$ a.u.), the cavity enhances the chemical rate. In this regime, $\kappa_\mathrm{PGH} \approx Y_\mathrm{cl}$ as $\kappa_\mathrm{GH} \approx 1$ (see red solid line in Fig.~\ref{fig:fig4}b). With increasing $\eta_\mathrm{c}$, the chemical rate is enhanced by up to $\approx 5$ times when $\eta_\mathrm{c} \approx 0.025$ a.u.  Further increase in $\eta_\mathrm{c}$ leads to a suppression of chemical kinetics (for $\eta_\mathrm{c} > 0.03$ a.u.) and   $\kappa_\mathrm{PGH} \approx \kappa_\mathrm{GH}$ as $Y_\mathrm{cl} \approx 1$ (see blue solid line in Fig.~\ref{fig:fig4}b). Overall, $\eta_\mathrm{c}$ plays a similar role to $\Lambda$ in modifying chemical kinetics, which is apparent when comparing Fig.~\ref{fig:fig1}b and Fig.~\ref{fig:fig4}b. It is worth noting, however, that  $\eta_\mathrm{c}$ can be tuned to reach enhancement {\it beyond} the maximum chemical rate possible by tuning $\Lambda$ alone in the bare molecular system (i.e. the peak of $\kappa_\mathrm{PGH}$ in Fig.~\ref{fig:fig1}b, also indicated by solid black line in Fig.~\ref{fig:fig4}b).

\begin{figure}
    \centering
    \includegraphics[width=0.9\columnwidth]{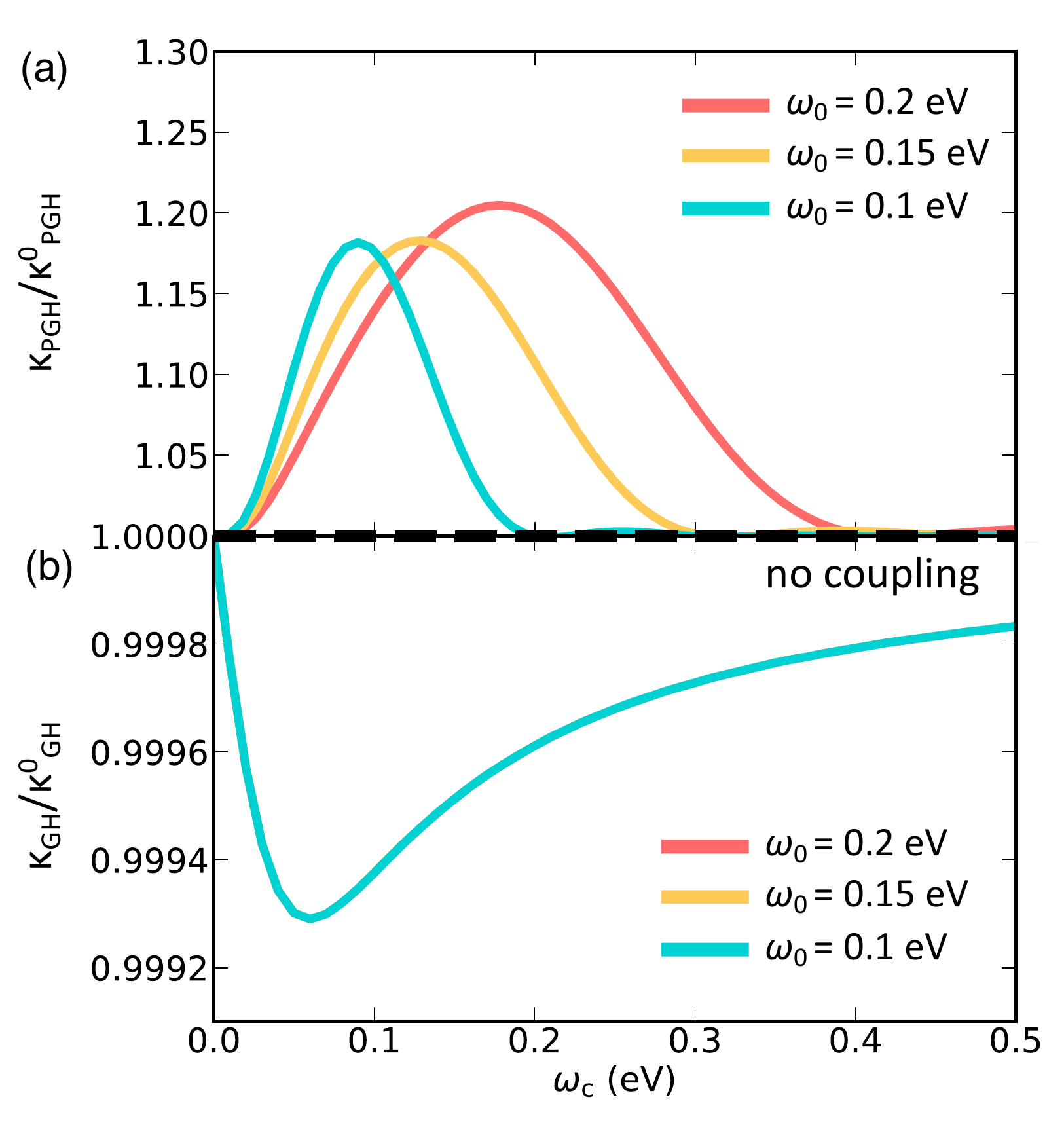}
    \caption{Reaction coordinate frequency dependence of the (a) Pollak-Grabert-H\"anggi (PGH) and (b) Grote-Hynes (GH)   transmission coefficients (normalized by the transmission coefficient in the absence of a cavity) in the energy-diffusion limited regime ($\Lambda = 2\times10^{-7}$ a.u. , $\eta = 0.0025$ a.u. and for varying reaction coordinate frequencies.}
    \label{fig:fig5}
\end{figure}

Finally, in Fig.~\ref{fig:fig5} we show the  $\omega_{0}$ (reactant well frequency) dependence of the cavity-modified chemical rate. As mentioned before, the depopulation factor $Y_\mathrm{cl}$ is peaked when $\omega_\mathrm{c}$ is close to $\omega_{0}$. This results in a cavity modification of the chemical rate that is peaked when $\omega_\mathrm{c}\approx \omega_\mathrm{0}$ (a ``resonance" effect) in the energy-diffusion limited regime. This is shown in Fig.~\ref{fig:fig5}a where we plot $\kappa_\mathrm{PGH}/\kappa_\mathrm{PGH}^{0}$ for three different values of $\omega_{0}$. Importantly, we observe that in addition to the maximum enhancement occurring at resonance, the rate profile becomes sharper with decreasing $\omega_{0}$. In comparison, the GH theory provides an $\omega_{0}$-independent rate profile as shown in Fig.~\ref{fig:fig5}b (all three curves are identical, such that the other two curves are hidden behind the green solid line). 

The $\omega_{0}$ dependence in the energy-diffusion limited regime can be leveraged to achieve mode selective chemistry in single molecule-cavity setups. This is in contrast to the spatial-diffusion limited regime, where selectivity can only be achieved when two chemical reactions have different barrier frequencies~\cite{LiJPCL2021}.

\section{Conclusions and outlook}\label{conclusion}
In conclusion, we have developed an analytical rate theory to describe cavity-modified ground state chemical kinetics at the single molecule level for the complete range of solvent friction. Our formulation is based on the Pollak-Grabert-H\"anggi (PGH) rate theory~\cite{PollakJCP1989} which includes a depopulation factor that is absent in previous works based on the Grote-Hynes (GH) theory~\cite{LiNC2021, LiJPCL2021, MandalJCP2022, YangJPCL2021}.  For weak solvent friction (the energy-diffusion limited regime), introduction of light-matter coupling leads to a sizable increase in the chemical reaction rate. We show that it is also possible to drive a molecular system initially in the energy-diffusion limited regime to the spatial-diffusion limited regime by increasing the light-matter coupling strength, as can be done by increasing solvent friction. When solvent friction places the system in the spatial-diffusion limited regime, increase in the light-matter coupling suppresses chemical reactivity. Here, the PGH theory reduces to the GH theory.  

Importantly we find that when the system is in the energy-diffusion limited regime, cavity coupling leads to a resonance effect, that is the cavity modification peaks when photon frequency is close to the reactant well frequency ($\omega_{0} \approx \omega_\mathrm{c}$). We show that this effect originates from the depopulation factor. In this case, the rate profile is much more dramatically altered and much sharper in comparison to the predictions of GH theory, and the width of the rate profile decreases with decreasing $\omega_{0}$.

There are several important limitation to the theory presented in this work as it pertains to direct comparison to recent experiments. First, our theory is completely classical. Quantum effects may become important when the cavity is in near-resonance with vibrational states within $V(Q)$. We are currently pursuing exact numerical calculation of the fully quantized Hamiltonian to include such effects. Secondly, our calculations only pertain to the strict single molecule limit (infinite dilution), and cannot address collective effects. A full quantum mechanical calculation of the case of $N$ molecules inside a cavity is prohibitively difficult. However, our fully quantum mechanical calculations performed in the single molecule case should enable the benchmarking of the accuracy of approximate quantum master equation approaches such as Redfield theory, which can open the door to the inclusion of both quantum and collective effects simultaenously. These important studies will be presented in future publications.  
%We should emphasize that while in this work we have considered a single reaction coordinate coupled to a cavity, it is not necessarily limited to a single molecule cavity setup. For example, the reaction coordinate could also represent a collective coordinate such as the Marcus coordinate in electron transfer~\cite{GargJCP1985}. In such scenario the light-matter coupling is not limited by the dipole of a single molecule. That said, collective effects observed in experiments remain unresolved and future theoretical works are needed to see if the present theory can be extended to ensemble of molecules coupled to cavity. 

\begin{acknowledgements}
This work was supported by NSF-1954791 (A.M. and D.R.R.) and by the Chemical Sciences, Geosciences, and  Biosciences Division of the Office of Basic Energy Sciences, Office of Science, U.S. Department of Energy (L.P.L. and D.R.R.). 
\\
\\
\\
\\
\end{acknowledgements}

\bibliography{main.bib}
\end{document}

\title{Supporting Information: Resonant Cavity Modification of Ground State Chemical Kinetics}
\author{Lachlan P. Lindoy}%
\affiliation{Department of Chemistry, Columbia University, 3000 Broadway, New York, New York, 10027,  U.S.A}
\author{Arkajit Mandal}%
\affiliation{Department of Chemistry, Columbia University, 3000 Broadway, New York, New York, 10027,  U.S.A}
\author{David R. Reichman}
%\email{drr2103@columbia.edu}
\affiliation{Department of Chemistry, Columbia University, 3000 Broadway, New York, New York, 10027,  U.S.A}

\maketitle
\section{Details of the molecular Hamiltonian} 
 Here we provide details for obtaining the molecular potential energy surface $V(Q)$ that is used in the main text. We define $V(Q)$ as the lowest eigenvalue of the following $3\times3$ matrix 
 \begin{align}
\begin{bmatrix}
E_{r}(Q) & \epsilon & 0\\
\epsilon & E_{b}(Q) & \epsilon\\
0 & \epsilon & E_{p}(Q) \\
\end{bmatrix},
 \end{align}
 where $\epsilon = 5$ meV, $E_{r}(Q) = \frac{1}{2}\omega_{0}(Q-Q_{r})^{2}$ and $E_{p}(Q) = \frac{1}{2}\omega_{0}(Q+Q_{r})^{2} $ are harmonic potentials with frequency $\omega_{0} = 0.2$ eV. $E_{b}(Q)$ is given as
  \begin{align}
  E_{b}(Q) &= E_{b} - \frac{1}{2}\omega_{b}^{2}Q^{2} \nonumber\\
  &+ \epsilon_{\infty} \Big(1-\Theta(Q-Q_{r}) + \Theta(Q+Q_{r})\Big)  ,
  \end{align}
where $\Theta(Q\mp Q_{r}) = \frac{1}{2}(1-\tanh(Q\mp Q_{r}))$ is a function centered around $\pm Q_{r}$, $\omega_{b} = 60$ meV is the barrier frequency and $E_{b} = 370$ meV is the barrier height.  Here we have used $Q_{r} = 40$ a.u. when using $\omega_{0} = 0.2$ eV. In Fig.~5, for $\omega_{0} = 0.15$ eV and $\omega_{0} = 0.1$ eV we have used $Q_{r} = 50$ a.u. and  $60$ a.u., respectively.

\section{Evaluation of the Unstable Mode Force}
Here we provide details for the numerical approach  to  obtain the unstable mode force, $F(t)$, used in the main text.  To perform this calculation, it is necessary to obtain the unstable mode trajectory, $\mathcal{Q}(t)$.  We introduce the variable $\mathcal{P}(t) = \dot{\mathcal{Q}}(t)$ and consider the set of first order ordinary-differential equations,
\begin{align}
\dot{\mathcal{Q}}(t) = \mathcal{P}(t), &&
\dot{\mathcal{P}}(t) = \lambda_0^2\mathcal{Q}(t) + F(t),
\end{align}
where $F(t)$ is defined in Eq. 15 of the main text.  This system of first-order ordinary differential equations are solved numerically, using standard numerical integration techniques, subject to the initial conditions
\begin{align}
    \mathcal{Q}(0) = \mathcal{Q}_b, && \mathcal{P}(0) = \mathcal{P}_0.
\end{align}
Here, $\mathcal{Q}_b$ is the barrier position, and $\mathcal{P}_0$ is a convergence parameter that may be either positive or negative depending on whether we are interested in the unstable mode trajectory associated with the right or left well, respectively.  These equations of motion are solved until, after reaching a turning point, the trajectory returns to the barrier position at some later time, $\tau$, i.e. $\mathcal{Q}(\tau) = \mathcal{Q}_b$.  This trajectory is defined over times $t=\left[0, \tau\right]$.  We obtain the unstable mode trajectory, as defined in the main text, by shifting the time domain to  $t=\left[-\tau/2, \tau/2\right]$. The unstable mode force can then be computed using Eq. 15 from the main text.  This process is repeated with decreasing values of $\|\mathcal{P}_0\|$, which results in increasing values of $\tau$, until the Fourier transform of $F(t)$ is converged to within some specified tolerance.
%\bibliography{vsc.bib}
%\bibliography{main.bib}